\documentstyle[sprocl,psfig]{article}

\def\NPB{{\em Nucl. Phys.} B}
\def\PLB{{\em Phys. Lett.}  B}
\def\PRL{\em Phys. Rev. Lett.}
\def\PRD{{\em Phys. Rev.} D}
\def\ZPC{{\em Z. Phys.} C}

\def\be{\begin{equation}}
\def\ee{\end{equation}}
\def\bea{\begin{eqnarray}}
\def\eea{\end{eqnarray}}
\def\be{\begin{equation}}
\def\ee{\end{equation}}
\def\bea{\begin{eqnarray}}
\def\eea{\end{eqnarray}}
\def\simlt{\stackrel{<}{{}_\sim}}
\def\simgt{\stackrel{>}{{}_\sim}}

\def\NPB#1#2#3{{\it Nucl.~Phys.} {\bf{B#1}} (19#2) #3}
\def\PLB#1#2#3{{\it Phys.~Lett.} {\bf{B#1}} (19#2) #3}
\def\PRD#1#2#3{{\it Phys.~Rev.} {\bf{D#1}} (19#2) #3}
\def\PRL#1#2#3{{\it Phys.~Rev.~Lett.} {\bf{#1}} (19#2) #3}
\def\ZPC#1#2#3{{\it Z.~Phys.} {\bf C#1} (19#2) #3}
\def\PTP#1#2#3{{\it Prog.~Theor.~Phys.} {\bf#1}  (19#2) #3}

\def\PR#1#2#3{{\it Phys.~Rep.} {\bf#1} (19#2) #3}
\def\RMP#1#2#3{{\it Rev.~Mod.~Phys.} {\bf#1} (19#2) #3}
\def\HPA#1#2#3{{\it Helv.~Phys.~Acta} {\bf#1} (19#2) #3}

\begin{document}

\begin{flushright}
IEM-FT-115/95 \\
hep--ph/9509395 \\
\end{flushright}

\vspace{1cm}

\title{BOUNDS ON THE HIGGS MASS IN THE STANDARD MODEL AND THE
MINIMAL SUPERSYMMETRIC STANDARD MODEL
\footnote{Based on talk given at the {\it International Workshop
On Elementary Particle Physics: Present and Future},
Valencia, June 5 to 9, 1995. }  }

\author{ M. QUIROS}

\address{CERN, TH Division, CH-1211 Geneva 23, Switzerland\\
Instituto de Estructura de la Materia, Serrano 123,\\
28006-Madrid, Spain}


\maketitle\abstracts{
\begin{center}
{\bf Abstract}
\end{center}
Depending on the Higgs-boson and  top-quark masses,
$M_H$ and $M_t$, the effective potential of the {\bf Standard
Model} can develop a non-standard minimum for values of
the field much larger than the weak scale. In those cases the
standard minimum becomes metastable and the
possibility of decay to the non-standard one arises.  Comparison of
the decay rate to the non-standard minimum at finite (and
zero) temperature with the corresponding expansion rate
of the Universe allows to identify the region, in the ($M_H$,
$M_t$) plane, where the Higgs field is sitting at the standard
electroweak minimum.
In the {\bf Minimal Supersymmetric Standard Model},
approximate analytical expressions for the Higgs mass spectrum
and couplings are worked out, providing an
excellent approximation to the numerical results which include all
next-to-leading-log corrections. An appropriate
treatment of squark decoupling allows to consider large values of
the stop and/or sbottom mixing parameters and thus fix a reliable upper
bound on the mass of the lightest CP-even Higgs boson mass.
The discovery of the Higgs boson at LEP~2 might
put an upper bound (below the Planck scale)
on the scale of new physics $\Lambda$ and eventually disentangle between
the Standard Model and the Minimal Supersymmetric Standard Model.}

\vspace{1cm}
\begin{flushleft}
IEM-FT-115/95\\
September 1995 \\
\end{flushleft}

\newpage
\section{Lower bounds on the Standard Model Higgs mass}

For particular values of the Higgs boson and top quark masses, $M_H$ and $M_t$,
the effective potential of the Standard Model (SM) develops a deep non-standard
minimum for values of the field $\phi \gg G_F^{-1/2}$~\cite{L}.
In that case the
standard electroweak (EW) minimum becomes metastable and might decay into
the non-standard one. This means that the SM might not
accomodate certain regions
of the plane ($M_H$,$M_t$), a fact
which can be intrinsically interesting as evidence for
new physics. Of course, the mere existence of the non-standard minimum,
and also the decay rate
of the standard one into it, depends on the scale $\Lambda$ up to which
we believe the SM results. In fact, one can identify $\Lambda$
with the scale of new physics.

\subsection{When the EW minimum becomes metastable?}
The preliminary question one should ask is: When the standard EW
minimum becomes
me\-ta\-sta\-ble, due to the appearance of a deep non-standard
minimum? This question was
addressed in past years~\cite{L} taking into account leading-log (LL) and
part of next-to-leading-log (NTLL) corrections.
More recently, calculations have incorporated all
NTLL
corrections~\cite{AI,CEQ}
resummed to all-loop by the renormalization group equations (RGE),
and considered pole masses for the top-quark and
the Higgs-boson.
{}From the requirement of a stable (not metastable) standard EW minimum
we obtain a lower bound on
the Higgs mass, as a function of the top mass, labelled by the values of
the SM cutoff (stability bounds). Our
result~\cite{CEQ} is lower than previous estimates by ${\cal O}$(10) GeV.

\begin{figure}[hbt]
\centerline{
\psfig{figure=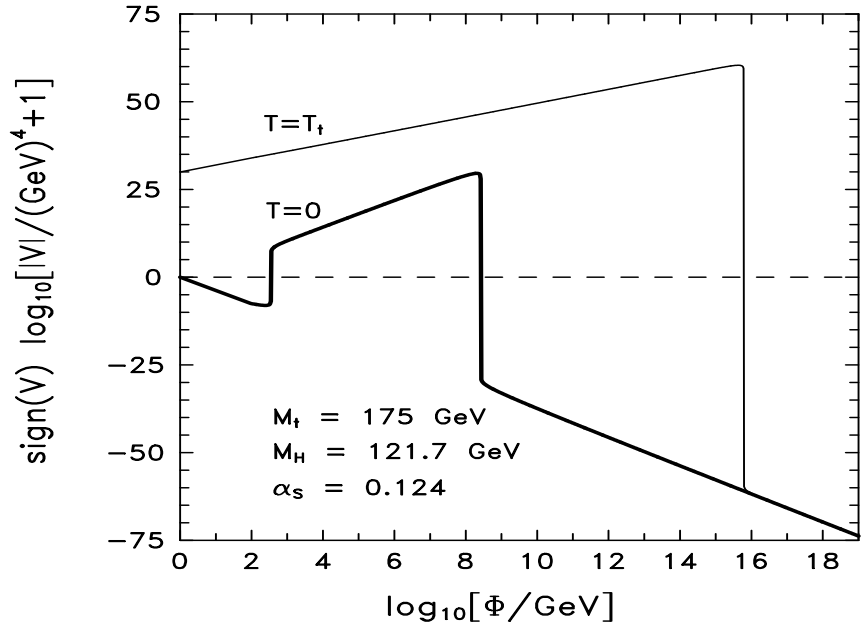,height=8.5cm,width=8cm,bbllx=4.75cm,bblly=3.cm,bburx=14.25cm,bbury=16cm}}
\caption{Plot of the effective potential for $M_t=175$ GeV, $M_H=121.7$
GeV at $T=0$ (thick solid line) and $T=T_t=2.5\times 10^{15}$ GeV
(thin solid line).}
\end{figure}
The one-loop effective potential of the SM improved by
two-loop RGE has been shown to
be highly scale independent~\cite{CEQR} and, therefore, very reliable for the
present study.
In Fig.~1 we show (thick solid line) the shape of the effective potential for
$M_t=175$ GeV
and $M_H=121.7$ GeV. We see the appearance of the non-standard maximum,
$\phi_M$, while the global
non-standard minimum has been cutoff at $M_P$. We can see from Fig.~1 the
steep descent from
the non-standard maximum. Hence, even if the non-standard minimum is beyond
the SM cutoff, the
standard minimum becomes metastable and can be destabilized. So for fixed
values of $M_H$ and
$M_t$ the condition for the standard minimum not to become metastable is
\be
\label{condstab}
\phi_M \simgt \Lambda
\ee
Condition (\ref{condstab}) makes the stability condition $\Lambda$-dependent.
In fact we have plotted
in Fig.~2 the stability condition on $M_H$ versus $M_t$ for
$\Lambda=
10^{19}$ GeV and 10 TeV. The stability
region corresponds to the region above the dashed curves.

\subsection{When the EW minimum decays?}

In the last subsection we have seen that in the region of Fig.~2
below the dashed line the standard EW minimum is
metastable. However we should not draw physical consequences
from this fact since we still do not
know at which minimum does the Higgs field sit. Thus, the real physical
constraint we have to impose is avoiding
the Higgs field sitting at its non-standard minimum.
In fact the Higgs field can be sitting at its
non-standard minimum at zero temperature because:
\begin{enumerate}
\item
The Higgs field was driven from the origin to the non-standard minimum
at finite temperature
by thermal fluctuations in a non-standard EW phase transition at
high temperature.
This minimum evolves naturally to the non-standard minimum at zero
temperature. In this case
the standard EW phase transition, at $T\sim 10^2$ GeV, will not take place.
\item
The Higgs field was driven from the origin to the
standard minimum at $T\sim 10^2$ GeV, but decays,
at zero temperature, to the non-standard minimum by a quantum fluctuation.
\end{enumerate}
In Fig.~1 we have depicted the effective potential at $T=2.5\times
10^{15}$ GeV (thin solid line) which
is the corresponding
transition temperature. Our finite temperature
potential~\cite{EQtemp} incorporates plasma effects~\cite{Q}
by one-loop resummation of Debye masses~\cite{DJW}.  The tunnelling
probability per unit time per
unit volume  was computed long ago for thermal~\cite{Linde} and
quantum~\cite{Coleman} fluctuations.
At finite temperature it is given by $\Gamma/\nu\sim T^4 \exp(-S_3/T)$,
where $S_3$ is the euclidean action evaluated
at the bounce solution $\phi_B(0)$. The semiclassical
picture is that unstable bubbles are nucleated behind the
barrier at $\phi_B(0)$ with a probability given by $\Gamma/\nu$. Whether
or not they fill the Universe depends on
the relation between the probability rate and the expansion
rate of the Universe. By normalizing the former
with respect to the latter we obtain a normalized probability $P$,
and the condition for decay corresponds
to $P\sim 1$. Of course our results are trustable,
and the decay actually happens, only if
$\phi_B(0)<\Lambda$, so that the similar condition to (\ref{condstab}) is
\be
\label{condmeta}
\Lambda< \phi_B(0)
\ee
\begin{figure}[htb]
\centerline{
\psfig{figure=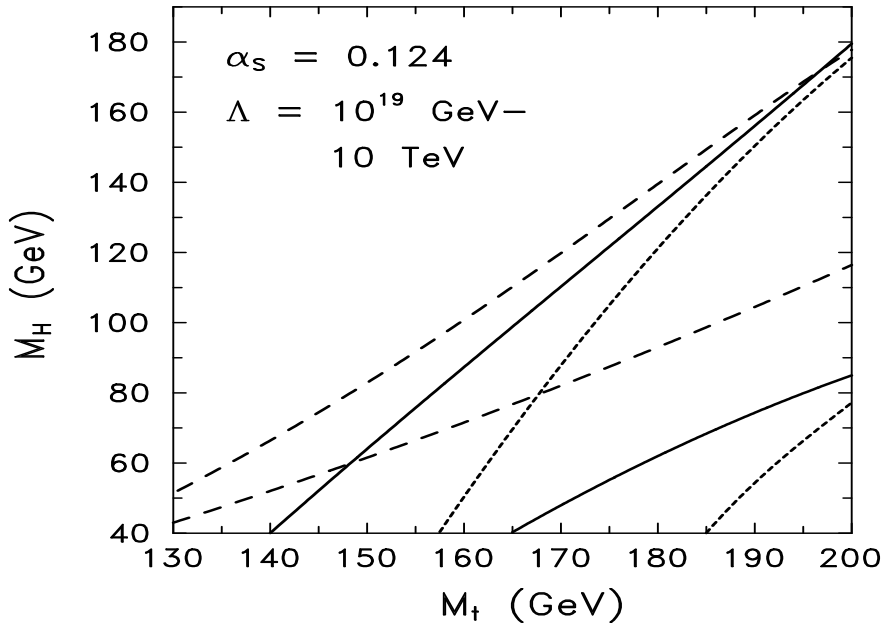,height=8.5cm,width=8cm,bbllx=5.cm,bblly=2.cm,bburx=14.5cm,bbury=15cm}}
\caption{Lower bounds on $M_H$ as a function of $M_t$, for
$\Lambda=10^{19}$ GeV (upper set) and $\Lambda=10$ TeV (lower set).
The dashed curves
correspond to the stability bounds of subsection 1.1 and  the solid (dotted)
ones to the metastability
bounds of subsection 1.2 at finite (zero) temperature.}
\end{figure}
The condition of no-decay (metastability condition) has
been plotted in Fig.~2  (solid lines)
for $\Lambda=10^{19}$ GeV and 10 TeV. The region
between the dashed and the solid line corresponds to
a situation where the non-standard minimum exists
but there is no decay to it at finite temperature.
In the region below the solid lines the Higgs field is sitting
already at the non-standard minimum at $T\sim 10^2$ GeV, and the  standard EW
phase transition does not happen.

We also have evaluated the tunnelling probability
at zero temperature from the standard EW minimum to the
non-standard one. The result of the calculation
should translate, as in the previous case, in lower bounds
on the Higgs mass for differentes
values of $\Lambda$. The corresponding bounds are shown in Fig.~2 in
dotted lines. Since the dotted lines
lie always below the solid ones, the possibility of quantum tunnelling at
zero temperature does not impose any extra constraint.

As a consequence of all improvements in the
calculation, our bounds are lower than previous
estimates~\cite{AV}. To fix ideas, for $M_t=175$ GeV, the
bound reduces by $\sim 10 $ GeV for $\Lambda=10^4$ GeV,
and $\sim 30$ GeV for $\Lambda=10^{19}$ GeV.

\section{Upper bounds on the Minimal Supersymmetrics Standard
Model lightest Higgs boson mass}

The {\bf effective potential} methods to compute the (radiatively
corrected) Higgs mass spectrum in the Minimal Supersymmetric
Standard Model (MSSM) are useful since they allow to {\bf resum}
(using Renormalization Group (RG) techniques) LL,
NTLL,..., corrections to {\bf all orders}
in perturbation theory. These methods~\cite{Effpot,EQ}, as well as the
diagrammatic methods~\cite{Diagram} to compute the Higgs mass spectrum
in the MSSM, were first developed in the early nineties.

Effective potential methods are based on the {\bf run-and-match}
procedure by which all dimensionful and dimensionless couplings
are running with the RG scale, for scales greater than the
masses involved in the theory. When the RG scale
equals a particular mass threshold, heavy fields decouple,
eventually leaving threshold effects in order to match the
effective theory below and above the mass threshold. For
instance, assuming a common soft supersymmetry breaking mass
for left-handed and right-handed stops and sbottoms,
$M_S\sim m_Q\sim m_U\sim m_D$, and assuming for the top-quark mass,
$m_t$, and for the CP-odd Higgs mass, $m_A$, the range
$m_t\leq m_A\leq M_S$, we have: for scales $Q\geq M_S$, the MSSM, for
$m_A\leq Q\leq M_S$ the two-Higgs doublet model (2HDM), and for
$m_t\leq Q\leq m_A$ the SM. Of course there are
thresholds effects at $Q=M_S$ to match the MSSM with the 2HDM, and
at $Q=m_A$ to match the 2HDM with the SM.

The neutral Higgs sector of the MSSM contains,
on top of the CP-odd Higgs $A$, two CP-even Higgs mass
eigenstates, $H_h$ (the heaviest one) and $H$ (the lightest one).
It turns out that the larger
$m_A$ the heavier the lightest Higgs $H$. Therefore the case
$m_A\sim M_S$ is, not only a great simplification since the effective
theory below $M_S$ is the SM, but also of great interest, since it
provides the upper bound on the mass of the lightest Higgs
(which is interesting for phenomenological purposes, e.g. at
LEP~2). In this case the threshold correction at $M_S$ for the SM
quartic coupling $\lambda$ is:
\be
\label{threshold}
\Delta_{\rm th}\lambda=\frac{3}{16\pi^2}h_t^4
\frac{X_t^2}{M_S^2}\left(2-\frac{1}{6}\frac{X_t^2}{M_S^2}\right)
\ee
where $h_t$ is the SM top Yukawa coupling and
$X_t=(A_t-\mu/\tan\beta)$ is the mixing in the stop mass
matrix, the parameters $A_t$ and $\mu$ being the trilinear
soft-breaking coupling in the stop sector and the supersymmetric
Higgs mixing mass, respectively. The maximum of
(\ref{threshold}) corresponds to $X_t^2=6 M_S^2$ which provides
the maximum value of the lightest Higgs mass: this case will be
referred to as the case of maximal mixing.
\begin{figure}[htb]
\centerline{
\psfig{figure=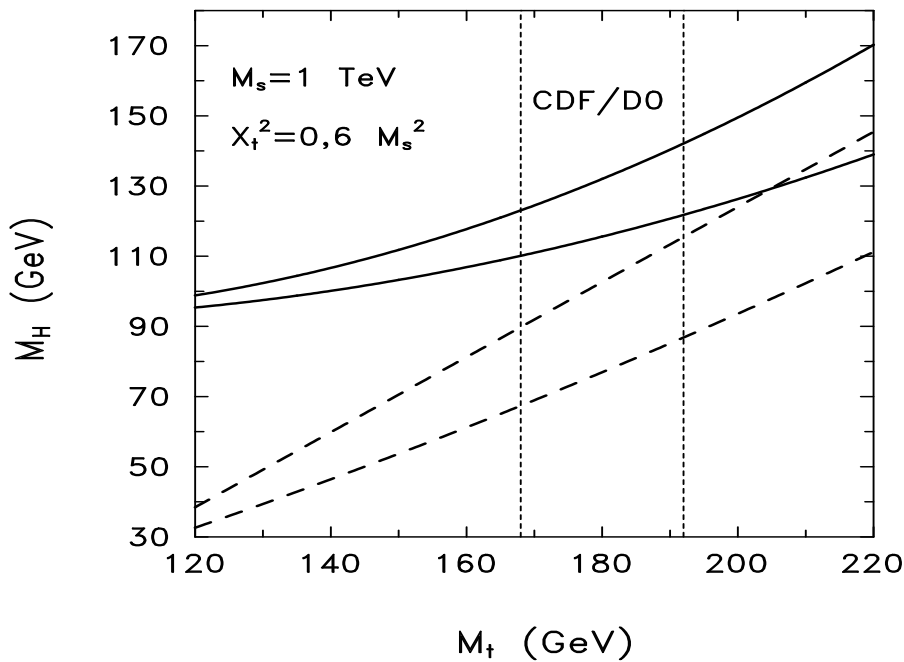,height=8.5cm,width=8cm,bbllx=5.5cm,bblly=2.5cm,bburx=15.cm,bbury=15.5cm}}
\caption{Plot of $M_H$ as a function of $M_t$ for $\tan\beta\gg 1$
(solid lines), $\tan\beta=1$ (dashed lines), and $X_t^2=6 M_S^2$ (upper set),
$X_t=0$ (lower set). The experimental band from the CDF/D0 detection is also
indicated.}
\end{figure}

We have plotted in Fig.~3 the lightest Higgs pole mass
$M_H$, where all NTLL corrections
are resummed to all-loop by the RG,
as a function of $M_t$~\cite{CEQR}. From Fig.~3 we
can see that the present experimental band from CDF/D0 for the
top-quark mass requires $M_H\simlt 140$ GeV, while if we fix
$M_t=170$ GeV, the upper bound $M_H\simlt 125$ GeV
follows. It goes without saying
that these figures are extremely relevant for MSSM Higgs searches
at LEP~2.

\subsection{An analytical approximation}

We have seen~\cite{CEQR} that,
since radiative corrections are minimized for scales $Q\sim m_t$,
when the LL RG improved Higgs mass expressions are
evaluated at the top-quark mass scale, they reproduce the NTLL value
with a high level of accuracy, for any value of $\tan\beta$ and the
stop mixing parameters~\cite{CEQW}
\be
\label{relmasas}
m_{H,LL}(Q^2\sim m_t^2)\sim m_{H,NTLL}.
\ee
Based on the above observation, we can work out a very accurate
analytical approximation to $m_{H,NTLL}$ by just keeping two-loop
LL corrections at $Q^2=m_t^2$, i.e. corrections of order $t^2$, where
$t=\log(M_S^2/m_t^2)$.

Again the case $m_A\sim M_S$ is the simplest, and very illustrative,
one. We have found~\cite{CEQW,HHH} that, in the absence of mixing
(the case $X_t=0$) two-loop corrections resum in the one-loop
result shifting the energy scale from $M_S$ (the tree-level scale)
to $\sqrt{M_S\; m_t}$. More explicitly,
\be
\label{resum}
m_H^2=M_Z^2 \cos^2 2\beta\left(1-\frac{3}{8\pi^2}h_t^2\; t\right)
+\frac{3}{2\pi^2 v^2}m_t^4(\sqrt{M_S m_t}) t
\ee
where $v=246.22$ GeV.

In the presence of mixing ($X_t\neq 0$), the run-and-match procedure
yields an extra piece in the SM effective potential
$\Delta V_{\rm th}[\phi(M_S)]$ whose second derivative gives an
extra contribution to the Higgs mass, as
\be
\label{Deltathm}
\Delta_{\rm th}m_H^2=\frac{\partial^2}{\partial\phi^2(t)}
\Delta V_{\rm th}[\phi(M_S)]=
\frac{1}{\xi^2(t)}
\frac{\partial^2}{\partial\phi^2(t)}
\Delta V_{\rm th}[\phi(M_S)]
\ee
which, in our case, reduces to
\be
\label{masthreshold}
\Delta_{\rm th}m_H^2=
\frac{3}{4\pi^2}\frac{m_t^4(M_S)}{v^2(m_t)}
\frac{X_t^2}{M_S^2}\left(2-\frac{1}{6}\frac{X_t^2}{M_S^2}\right)
\ee
%
\begin{figure}[ht]
\centerline{
\psfig{figure=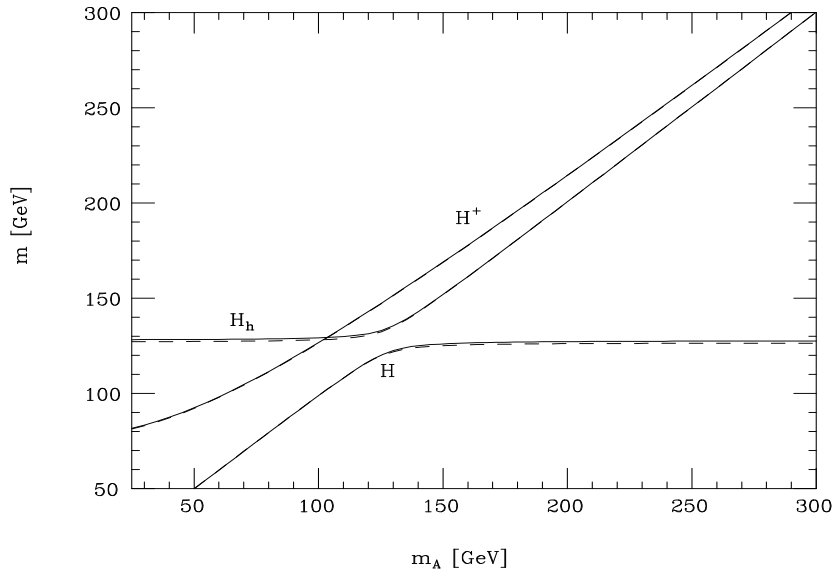,height=9.5cm,width=14cm,angle=90} }
\caption[0]
{The neutral ($H_h,H$) and charged ($H^+$) Higgs mass spectrum
as a function of the CP-odd Higgs mass $m_A$ for
a physical top-quark mass $M_t =$ 175 GeV and $M_S$ = 1 TeV, as
obtained from the one-loop improved RG evolution
(solid lines) and the analytical formulae (dashed lines).
All sets of curves correspond to
$\tan \beta=$ 15 and large squark mixing, $X_t^2 = 6 M_S^2$
($\mu=0$).}
\end{figure}

We have compared our analytical approximation~\cite{CEQW}
with the numerical NTLL result~\cite{CEQR} and found a difference
$\simlt 2$ GeV for all values of supersymmetric parameters.

The case $m_A<M_S$ is a bit more complicated since the effective theory
below the supersymmetric scale $M_S$ is the 2HDM. However since radiative
corrections in the 2HDM are equally dominated by the top-quark, we can
compute analytical expressions based upon the LL approximation
at the scale $Q^2\sim m_t^2$. Our approximation~\cite{CEQW} differs from
the LL all-loop numerical resummation by $\simlt 3$ GeV, which we
consider the uncertainty inherent in the theoretical calculation,
provided the mixing is moderate and, in particular, bounded by the
condition,
\be
\label{condicion}
\left|\frac{m^2_{\;\widetilde{t}_1}-m^2_{\;\widetilde{t}_2}}
{m^2_{\;\widetilde{t}_1}+m^2_{\;\widetilde{t}_2}}\right|\simlt 0.5
\ee
where $\widetilde{t}_{1,2}$ are the two stop mass eigenstates.
In Fig.~4 the Higgs mass spectrum is plotted versus $m_A$.

\subsection{Threshold effects}

There are two possible caveats in the
analytical approximation we have just
presented: {\bf i)} Our expansion parameter $\log(M_S^2/m_t^2)$
does not behave properly in the supersymmetric limit $M_S\rightarrow 0$,
where we should recover the tree-level result. {\bf ii)} We have expanded
the threshold function $\Delta V_{\rm th}[\phi(M_S)]$ to order $X_t^4$.
In fact keeping the whole threshold function $\Delta V_{\rm th}[\phi(M_S)]$
we would be able to go to larger values of $X_t$ and to evaluate the
accuracy of the approximation (\ref{threshold}) and (\ref{masthreshold}).
Only then we will be able to
check the reliability of the maximum value of the
lightest Higgs mass (which corresponds to the maximal mixing) as provided
in the previous sections.
This procedure has been properly followed~\cite{CEQW,CQW} for
the most general case $m_Q\neq m_U\neq m_D$.
We have proved that keeping the exact threshold function
$\Delta V_{\rm th}[\phi(M_S)]$, and properly running its value from the
high scale to $m_t$ with the corresponding anomalous dimensions as in
(\ref{Deltathm}), produces two effects: {\bf i)} It makes a resummation
from $M_S^2$ to $M_S^2+m_t^2$ and generates as (physical) expansion
parameter $\log[(M_S^2+m_t^2)/m_t^2]$. {\bf ii)} It generates a whole
threshold function $X_t^{\rm eff}$ such that (\ref{masthreshold})
becomes
\be
\label{masthreshold2}
\Delta_{\rm th}m_H^2=
\frac{3}{4\pi^2}\frac{m_t^4[M_S^2+m_t^2]}{v^2(m_t)}
X_t^{\rm eff}
\ee
and
\be
\label{desarrollo}
X_t^{\rm eff}=\frac{X_t^2}{M_S^2+m_t^2}
\left(2-\frac{1}{6}\frac{X_t^2}{M_S^2+m_t^2}\right)+\cdots
\ee
The numerical calculation shows~\cite{CQW} that $X_t^{\rm eff}$
has the maximum very close to $X_t^2=6(M_S^2+m_t^2)$,
what justifies
the reliability of previous upper bounds on the lightest Higgs mass.

\section{Would a light Higgs detection imply new physics?}

{}From the previous sections it should be clear by
now that the Higgs and top mass measurements
could serve to discriminate between the SM and its extensions,
and to provide information about the
scale of new physics $\Lambda$.
\begin{figure}[htb]
\centerline{
\psfig{figure=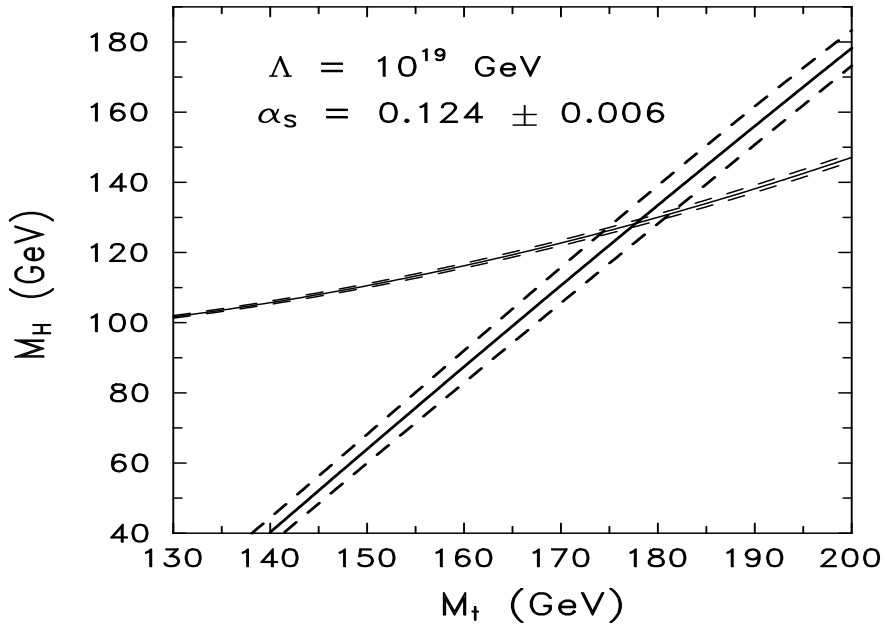,height=8.5cm,width=8cm,bbllx=5.cm,bblly=2.cm,bburx=14.5cm,bbury=15cm}}
\caption{SM lower bounds on $M_H$ (thick lines) as a function of
$M_t$, for $\Lambda=10^{19}$ GeV, from metastability requirements,
and upper bound on the lightest Higgs boson mass in the MSSM
(thin lines) for $M_S=1$ TeV.}
\end{figure}
In Fig.~5
we give the SM lower bounds on
$M_H$ for $\Lambda\simgt 10^{15}$ (thick lines) and
the upper bound on the mass of the
lightest Higgs boson in the MSSM (thin lines)
for $M_S\sim 1$ TeV. Taking,
for instance,  $M_t=180$ GeV, which coincides with the
central value recently reported by CDF+D0~\cite{top},
and $M_H\simgt 130$ GeV, the SM is
allowed and the MSSM is excluded. On the other hand,
if $M_H\simlt 130$ GeV, then the MSSM is
allowed while the SM is excluded. Likewise there
are regions where the SM is excluded, others
where the MSSM is excluded and others where both are permitted or
both are excluded.
\begin{figure}[htb]
\centerline{
\psfig{figure=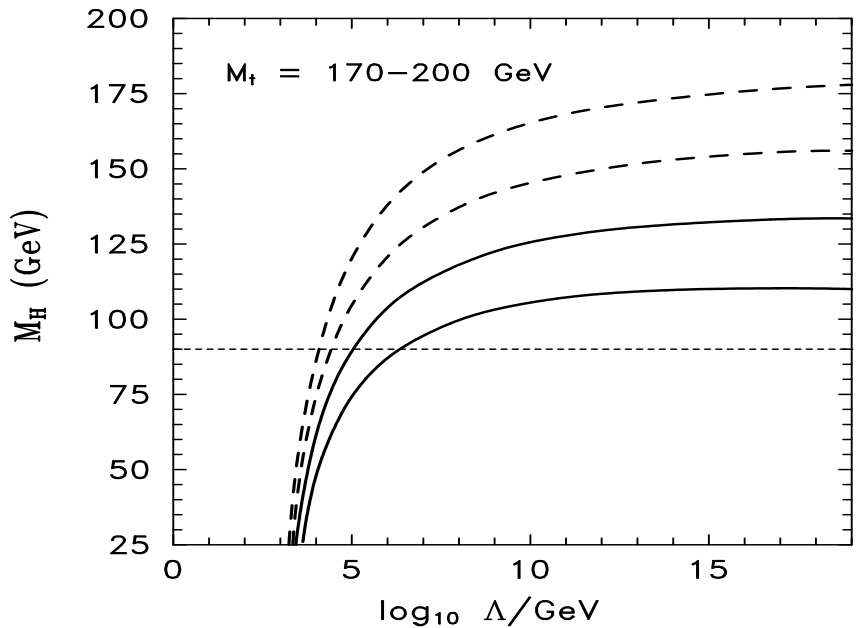,height=8.5cm,width=8cm,bbllx=5.cm,bblly=2.5cm,bburx=14.5cm,bbury=15.5cm}}
\caption{SM lower bounds on $M_H$ from
metastability requirements as a function of
$\Lambda$ for different values of $M_t$.}
\end{figure}

Finally from the bounds $M_H(\Lambda)$ (see Fig.~6)
one can easily deduce that
a measurement of $M_H$ might provide an
{\bf upper bound}  (below the Planck scale) on the
scale of new physics provided that
\be
\label{final}
M_t>\frac{M_H}{2.25\;  {\rm GeV}}+123\; {\rm GeV}
\ee
Thus, the present
experimental bound from LEP, $M_H>64$ GeV, would imply, from
(\ref{final}), $M_t>152$ GeV, which is fulfilled
by experimental detection of the
top~\cite{top}. Even non-observation of the Higgs at
LEP~2 (i.e. $M_H\simgt 95$ GeV), would
leave an open window ($M_t\simgt 163$ GeV)
to the possibility that a future Higgs detection
at LHC could lead to an upper bound on $\Lambda$. Moreover, Higgs detection at
LEP~2 would put an upper bound on the scale of new physics. Taking,
for instance,  $M_H\simlt 95$
GeV and  170 GeV $< M_t< $ 180 GeV, then $\Lambda\simlt 10^7$ GeV, while for
180 GeV $< M_t <$ 190 GeV, then $\Lambda\simlt 10^4$
GeV, as can be deduced from Fig.~6.

\section*{Acknowledgments}

Work supported in part by
the European Union (contract CHRX-CT92-0004) and
CICYT of Spain (contract AEN94-0928).

\end{document}